\documentclass{aastex}
\usepackage{spr-astr-addons}
\usepackage{url}

\RequirePackage{color}

\begin{document}

\title{Interacting modified Chaplygin gas in loop quantum cosmology}

\shorttitle{Chaplygin gas in quantum cosmology} \shortauthors{Jamil
et al.}

\author{Mubasher Jamil\altaffilmark{1}}
\and
\author{Ujjal Debnath\altaffilmark{2}}

\altaffiltext{1}{Center for Advanced Mathematics and Physics,
National University of Sciences and Technology, H-12, Islamabad,
Pakistan. Email: mjamil@camp.nust.edu.pk , jamil.camp@gmail.com}

\altaffiltext{2}{Department of Mathematics, Bengal Engineering and
Science University, Shibpur, Howrah-711 103, India. Email:
ujjaldebnath@yahoo.com , ujjal@iucaa.ernet.in}

\begin{abstract}
We investigate the background dynamics when dark energy is coupled
to dark matter in the universe described by loop quantum cosmology.
We consider dark energy of the form modified Chaplygin gas. The
dynamical system of equations is solved numerically and a stable
scaling solution is obtained. It henceforth resolves the famous
cosmic coincidence problem in modern cosmology. The statefinder
parameters are also calculated to classify this dark energy model.
\end{abstract}

\section{Introduction}

Recent observations of type Ia Supernovae indicate that Universe is
expanding with acceleration \citep{perl,ries} and lead to the search
for a new type of matter which violates the strong energy condition,
i.e., $\rho+3p<0$. In Einstein's general relativity, an energy
component with large negative pressure has to be introduced in the
total energy density of the Universe in order to explain this cosmic
acceleration. This energy component is known as {\it dark energy}
\citep{sahni,paddy}. There are many candidates supporting this
behavior \citep{cope}, scalar field or quintessence \citep{pee}
being one of the most favored candidates as it has a decaying
potential term which dominates over the kinetic term thus generating
enough pressure to drive acceleration.

Presently we live in an epoch where the densities of the dark energy
and the dark matter are comparable. It becomes difficult to solve
this coincidence problem without a suitable interaction. Generally
interacting dark energy models are studied to explain the cosmic
coincidence problem \citep{jamil1,jamil2,jamil3,jamil4,jamil5}. Also
the transition from matter domination to dark energy domination can
be explained through an appropriate energy exchange rate. Therefore,
to obtain a suitable evolution of the Universe an interaction is
assumed and the decay rate should be proportional to the present
value of the Hubble parameter for good fit to the expansion history
of the Universe as determined by the Supernovae and CMB data
\citep{jamil1,jamil2}. A variety of interacting dark energy models
have been proposed and studied for this purpose
\citep{setare1,setare2,hu,wu,jamil6,setare67}.

In recent years, the model of interacting dark energy has been
explored in the framework of loop quantum cosmology (LQC) as well:
It is shown in \citep{wu1} that for the quintessence model, the
cosmological evolution in LQC is the same as that in classical
Einstein cosmology, whereas for the phantom dark energy the loop
quantum effect significantly reduce the parameter spacetime required
by stability. In \citep{chen}, the authors used a more general
interaction term to study the interacting dark energy. They showed
that in LQC, the parameter space for the existence of the
accelerated scaling attractor is found to be smaller then that in
Einstein cosmology. In another study \citep{fu}, the authors studied
the model with an interacting phantom scalar field with an
exponential potential and deduced that the future singularity
appearing in the standard FRW cosmology can be avoided by loop
quantum effects.

In this paper, we extend the model of interacting modified Chaplygin
gas (MCG) from the framework of Einstein gravity to LQC. We
construct a dynamical system of equations and solve them
numerically. We obtain a stable scaling solution (which is also an
`attractor') of modified FRW equations. We discuss our results in
the final section.

\section{The model}

The modified Friedmann equation for LQC is given by
\citep{wu1,chen,fu}
\begin{equation}\label{1}
H^2=\frac{\rho}{3}\Big( 1-\frac{\rho}{\rho_1} \Big).
\end{equation}
Here $\rho_1\equiv\sqrt{3}\pi^2\gamma^3G^2\hbar$ is the critical
loop quantum density and $\gamma$ is the dimensionless
Barbero-Immirzi parameter. We assume the interaction between dark
energy and pressureless dark matter. Hence the energy balance
equations for the interacting dark energy and dark matter can be
expressed as
\begin{eqnarray}
\dot{\rho}_\text{mcg}+3H(1+\omega_\text{mcg})\rho_\text{mcg}&=&-Q,\\\label{2}
\dot{\rho}_\text{m}+3H\rho_\text{m}&=&Q,\label{3}
\end{eqnarray}
where $Q=3bH\rho$ is the interaction term, $b$ is the coupling
parameter (or transfer strength) and
$\rho=\rho_\text{mcg}+\rho_\text{m}$ is the total cosmic energy
density which satisfies $\dot{\rho}+3H(\rho+p)=0$
\citep{guo,campo22}. Note that addition of the above two equations
leads to the energy conservation. Due to unknown nature of both dark
energy and dark matter, the interaction term can not be derived from
the first principles. It is worthy to note that if $Q<0$ than it
will yield the energy density of dark energy to be negative at
sufficiently early times, consequently the second law of
thermodynamics can be violated \citep{lima} hence $Q$ must be
positive and small. From the observational data of 182 Gold type Ia
supernova samples, CMB data from the three year WMAP survey and the
baryonic acoustic oscillations from the Sloan Digital Sky Survey, it
is estimated that the coupling parameter between dark matter and
dark energy must be a small positive value (of the order unity),
which satisfies the requirement for solving the cosmic coincidence
problem and the second law of thermodynamics \citep{feng}. Because
of the underlying interaction, the beginning of the accelerated
expansion is shifted to higher redshifts.

Consequently we obtain the modified Raychaudhuri equation
\begin{equation}\label{4}
\dot H=-\frac{1}{2}(\rho+p)\Big( 1-2\frac{\rho}{\rho_1} \Big),
\end{equation}
where $p$ is the total pressure ($p=p_\text{mcg}$). We shall use
modified Chaplygin gas as the dark energy. The MCG equation of state
is given by
\begin{equation}\label{4a}
p_\text{mcg}=A\rho_\text{mcg}-\frac{B}{\rho_\text{mcg}^\alpha},
\end{equation}
where $A$, $B$ and $\alpha$ are constants. The MCG best fits with
the $3-$year WMAP and the SDSS data with the choice of parameters
$A=-0.085$ and $\alpha=1.724$ \citep{lu} which are improved
constraints than the previous ones $-0.35<A<0.025$ \citep{jun}.
Recently it is shown that the dynamical attractor for the MCG exists
at $\omega_\text{mcg}=-1$, hence MCG crosses this value from either
side $\omega_\text{mcg}>-1$ or $\omega_\text{mcg}<-1$, independent
to the choice of model parameters \citep{jing}. A generalization of
MCG is suggested in \citep{debnath} by considering $B\equiv B(a)=B_o
a^{n}$, where $n$ and $B_o$ are constants. The MCG is the
generalization of generalized Chaplygin gas
$p_\text{mcg}=-B/\rho_\text{mcg}^\alpha$ \citep{sen,carturan} with
the addition of a barotropic term. This special form also appears to
be consistent with the WMAP $5-$year data and henceforth the support
the unified model with dark energy and matter based on generalized
Chaplygin gas \citep{barriero,makler,setarea,setareb,setarec}. In
the cosmological context, the Chaplygin gas was first suggested as
an alternative to quintessence and demonstrated an increasing
$\Lambda$ behavior for the evolution of the universe
\citep{kamenshchik}. Recent supernovae data also favors the
two-fluid cosmological model with Chaplygin gas and matter
\citep{grigoris}.

To analyze the dynamical system, we convert the physical parameters
into dimensionless form as
\begin{equation}\label{5}
x=\ln a,\ \ u=\frac{\rho_\text{mcg}}{3H^2},\ \
v=\frac{\rho_\text{m}}{3H^2},
\end{equation}
where $a_0=1$ is assumed, where the subscript 0 refers to the
present time. Making use of (\ref{1}) to (\ref{5}), we can write
\begin{eqnarray}\label{7}
\frac{du}{dx}&=&-3b(u+v)-3u(1+\omega_\text{mcg})\nonumber\\&&+3u[u(1+
\omega_\text{mcg})+v]
\Big( -1+\frac{2}{u+v} \Big),\\
\frac{dv}{dx}&=&3b(u+v)-3v+3v[u(1+\omega_\text{mcg})+v]\nonumber\\&&\times
\Big( -1+\frac{2}{u+v} \Big),\label{6}
\end{eqnarray}
where the state parameter of modified Chaplygin gas is
\begin{equation}
\omega_\text{mcg}=\frac{p_\text{mcg}}{\rho_\text{mcg}}=A-\frac{B(u+v)^{2(\alpha+
1)}}{\rho_{1}^{\alpha+1}u^{\alpha+1}(u+v-1)^{\alpha+1}}.
\end{equation}
For the mathematical simplicity, we work out $\alpha=1$ only. The
critical points of the above system are obtained by putting
$\frac{du}{dx}=0=\frac{dv}{dx}$ which yield
\begin{eqnarray}\label{8}
u_{1c}&=&\Big(\frac{1}{1-b}+\frac{\sqrt{B}}{\sqrt{(-1+A(-1+b))
(-1+b)^{3}\rho_{1}^{2}
 }} \Big)^{-1},\nonumber \\
 v_{1c}&=&\frac{b\rho_{1}}{-B+(-1+A(-1+b))(-1+b)\rho_{1}^{2}}\nonumber\\&&
\times \Big[(-1+A(-1+b))(-1+b)\rho_{1}
 \nonumber\\&&+ \sqrt{B(-1+A(-1+b))(-1+b)
 }\Big],\\
u_{2c}&=&\Big(\frac{1}{1-b}-\frac{\sqrt{B}}{\sqrt{(-1+A(-1+b))
(-1+b)^{3}\rho_{1}^{2}
 }} \Big)^{-1},\nonumber \\
v_{2c}&=&\frac{b\rho_{1}}{B-(-1+A(-1+b))(-1+b)\rho_{1}^{2}}\nonumber\\&&
\times \Big[(-1+A(-1+b))(-1+b)\rho_{1}
 \nonumber\\&&+ \sqrt{B(-1+A(-1+b))(-1+b)
 }\Big].\label{9}
\end{eqnarray}
The two critical points correspond to the era dominated by dark
matter and MCG type dark energy and exist for $A>\frac{1}{b-1}$.

For the two critical points, the state parameter (9) of the
interacting dark energy takes the form
\begin{equation*}
w^i_\text{mcg}=A-\frac{B(u_{ic}+v_{ic})^{2(\alpha+1)}}{\rho_{c}^{\alpha+1}u_{ic}^{\alpha+1}
(u_{ic}+v_{ic}-1)^{\alpha+1}},\ \ \ i=1,2
\end{equation*}
which holds only when $u_{ic}+v_{ic}\neq1$.

We further check the stability of the dynamical system (Eqs. (7) and
(8)) about the critical point. To do this, we linearize the
governing equations about the critical point i.e. $u=u_c+\delta u$
and $v=v_c+\delta v$, we obtain
\begin{eqnarray}\label{10}
\delta\Big(\frac{du}{dx}\Big)&=&\Big[-3(-1+b+2u+v)\nonumber\\&&+A\Big\{-3+6u
\Big(-1+\frac{u+2v}{(u+v)^{2}} \Big) \Big\}
\nonumber\\&&+\frac{3B(u+v)^{2}}{u^{2}(-1+u+v)^{3}\rho_{1}^{2}
}\nonumber\\&&\times\Big\{2u^{4}-(-1+v)v^{2}-uv(2+v)\nonumber\\&&+u^{2}(-1+v)(-3+2v)
+u^{3}(-5+4v) \Big\} \Big]_c\delta u \nonumber\\&&
+\Big[-3b-3u\Big(1+\frac{2Au}{(u+v)^{2}} \Big)\nonumber\\&&+
\frac{6B(u+v)^{2} }{u(-1+u+v)^{3}\rho_{1}^{2}
}\Big\{u^{3}+2u^{2}(-1+v)\nonumber\\&&+(-2+v)v+u(1+(-1+v)v) \Big\}
\Big]_c \delta v,\nonumber
\end{eqnarray}
\begin{eqnarray}\label{11}
\delta\Big(\frac{dv}{dx}\Big)&=&\Big[3b-3v\Big(1+\frac{A(-2v+(u+v)^{2})}{(u+v)^{2}}
\Big)\nonumber\\&&
+\frac{3Bv(u+v)^{2}}{u^{2}(-1+u+v)^{3}\rho_{1}^{2}
}\Big\{u^{3}+u^{2}(-3+v)\nonumber\\&&-(-2+v)(-1+v)v-u(-4+v^{2})
\Big\} \Big]_c\delta u \nonumber\\&&
+\Big[3b+\frac{3}{u(-1+u+v)^{3}(u+v)^{2}\rho_{1}^{2}}\nonumber\\&&\times
\Big\{ B(u+v)^{4}(u^{3} +u^{2}(-3+5v) \nonumber\\&&
+v(8+3(-3+v)v)+u(2+v(-12+7v)))\nonumber\\&&-u(-1+u+v)^{3}(u^{2}(-1+A(-2+u)+u)
\nonumber\\&&
+2u(-1+(2+A)u)v+(-1+(5+A)u)v^{2}\nonumber\\&&+2v^{3})\rho_{1}^{2}
\Big\}\Big]_c\delta v .\nonumber
\end{eqnarray}
The subscript $c$ refers to quantities evaluated at the critical
point of the dynamical system. We also calculate the deceleration
parameter $q=-1-(\dot H/H^2)$, in this model as
\begin{equation}\label{14}
q=-1+\frac{3}{2}\Big(1+\omega_\text{mcg}\frac{\rho_\text{mcg}}{\rho}\Big)
\Big(\frac{1-2\rho/\rho_1}{1-\rho/\rho_1} \Big),
\end{equation}
which can be written in terms of dimensionless density parameter
$\Omega_\text{mcg}=\rho_\text{mcg}/\rho$:
\begin{equation*}
q_\text{LQC}=-1+\frac{3}{2}\Big(1+\omega_\text{mcg}\Omega_\text{mcg}\Big)
\Big(\frac{1-2\rho/\rho_1}{1-\rho/\rho_1} \Big).
\end{equation*}
Clearly in the limit of $\rho_1\rightarrow\infty$, we retrieve the
result for the Einstein's gravity as
\begin{equation*}
q_\text{EG}=-1+\frac{3}{2}\Big(1+\omega_\text{mcg}\Omega_\text{mcg}\Big).
\end{equation*}
 Assuming
$\rho/\rho_1=\epsilon\sim O(1)$ and using (\ref{5}), we obtain
\begin{equation}\label{15}
q=-1+\frac{3}{2}\Big(1+\frac{u\omega_\text{mcg}}{u+v}\Big)\Big(\frac{1-2\epsilon}{1-\epsilon}
\Big).
\end{equation}
Since the only physically acceptable solution corresponds to the
first stable critical point, such that
$(u,v)\rightarrow(u_{1c},v_{1c})$. Hence (\ref{15}) gives
\begin{equation}\label{16}
q_c=-1+\frac{3}{2}X,\ \ \
X=\Big(1+\frac{\omega_\text{mcg}u_{1c}}{u_{1c}+v_{1c}}\Big)\Big(\frac{1-2\epsilon}{1-\epsilon}
\Big).
\end{equation}
As special cases, observe that for $\epsilon=1/2$, we have $q=-1$
while $\epsilon=1$ yields $q\rightarrow-\infty$ if
$\omega_\text{mcg}\geq-1/u_{1c}$.

Moreover the Hubble parameter varies as
\begin{equation}\label{17}
H=\frac{2}{3Xt},
\end{equation}
where we have ignored the integration constant. Integration of
(\ref{17}) yields
\begin{equation}\label{18}
a(t)=a_0t^{\frac{2}{3X}},
\end{equation}
which gives a power law form of the expansion.

We also calculate the statefinder parameters. Sahni et al
\citep{sahni1} introduced a pair of cosmological diagnostic pair
$\{r,s\}$ which they termed as Statefinder. The two parameters are
dimensionless and are geometrical since they are derived from the
cosmic scale factor alone, though one can rewrite them in terms of
the parameters of dark energy and matter. Additionally, the pair
gives information about dark energy in a model independent way i.e.
it categorizes dark energy in the context of background geometry
only which is not dependent on the theory of gravity. Hence
geometrical variables are universal. Also this pair generalizes the
well-known geometrical parameters like the Hubble parameter and the
deceleration parameter. This pair is algebraically related to the
equation of state of dark energy and its first time derivative.
\begin{equation}\label{19}
r\equiv\frac{\stackrel{...}a}{aH^3},\ \ s\equiv\frac{r-1}{3(q-1/2)}.
\end{equation}
In the present model, (\ref{19}) gives
\begin{eqnarray}\label{20}
r_\text{LQC}&=&\left(1-\frac{3X}{2}\right)(1-3X),\\
 s_\text{LQC}&=&2X.\label{21}
\end{eqnarray}
It is interesting to note that the pair
$\{r_\text{LQC},s_\text{LQC}\}$ yields the $\Lambda$CDM
(cosmological constant-cold dark matter model)
$\{r_\text{EG},s_\text{EG}\}=\{1,0\}$ when $X=0$ (or
$\epsilon=1/2$).

\section{Discussion}

In this work, we considered modified Friedmann model in loop quantum
cosmology. We assumed dark energy of the form modified Chaplygin
gas. The interaction between dark matter and MCG has been
investigated in LQC. The dynamical system of equations is solved
numerically and a stable scaling solution is obtained. It henceforth
resolves the famous cosmic coincidence problem in modern cosmology.
The deceleration parameter and statefinder parameters are also
calculated to classify this dark energy model. The dimensionally
density parameters $v$ and $u$ are drawn in figures 1 and 2. We see
that $v$ decreases and $u$ increases during evolution of the
universe. From figure 4, we also see that the ratio of the above
parameters decreases during time. The phase space diagram (figure 3)
shows the attractor solution hence the present state and the future
evolution of the universe is independent to the choice of initial
conditions. Moreover the expansion of the universe is governed by a
power-law form, rather than exponential or oscillatory. Hence the
expansion will go on forever with an ever increasing rate. The
variations of $q_\text{LQC}$, $r_\text{LQC}$ and $s_\text{LQC}$ are
shown in figure 5-7 respectively against $\omega_\text{mcg}$. It is
observed that the more negative the state-parameter of MCG, the more
negative values will be taken by the deceleration parameter. Finally
our results also reduce to those for Einstein's gravity under
suitable limits of parameters.

\section*{Acknowledgment} One of the authors (MJ) would like to
thank the Abdus Salam International Center for Theoretical Physics
(ICTP), Trieste, Italy where part of this work was completed. The
authors would also thank the referee for his useful criticism on
this work.

\begin{figure}
\includegraphics[scale=.4]{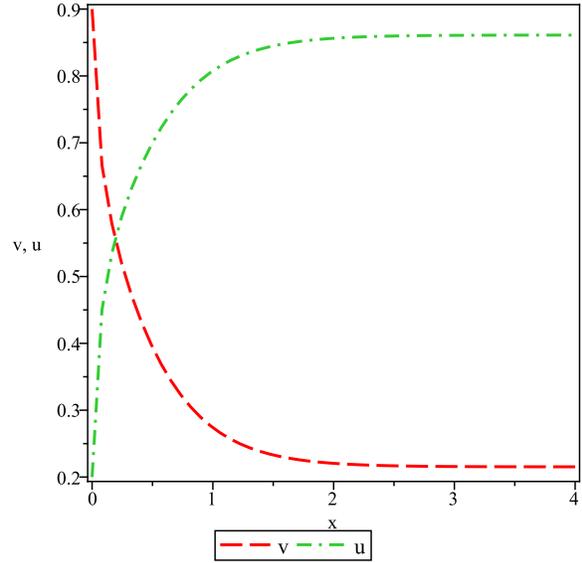}\\
\caption{The dimensionless density parameters are plotted against
e-folding time. The initial condition is $v(0)=0.9$, $u(0)=0.2$.
Other parameters are fixed at $b=0.2$, $A=0.3$, $B=0.5$ and
$\alpha=1$.}
\end{figure}
\begin{figure}
\includegraphics[scale=.4]{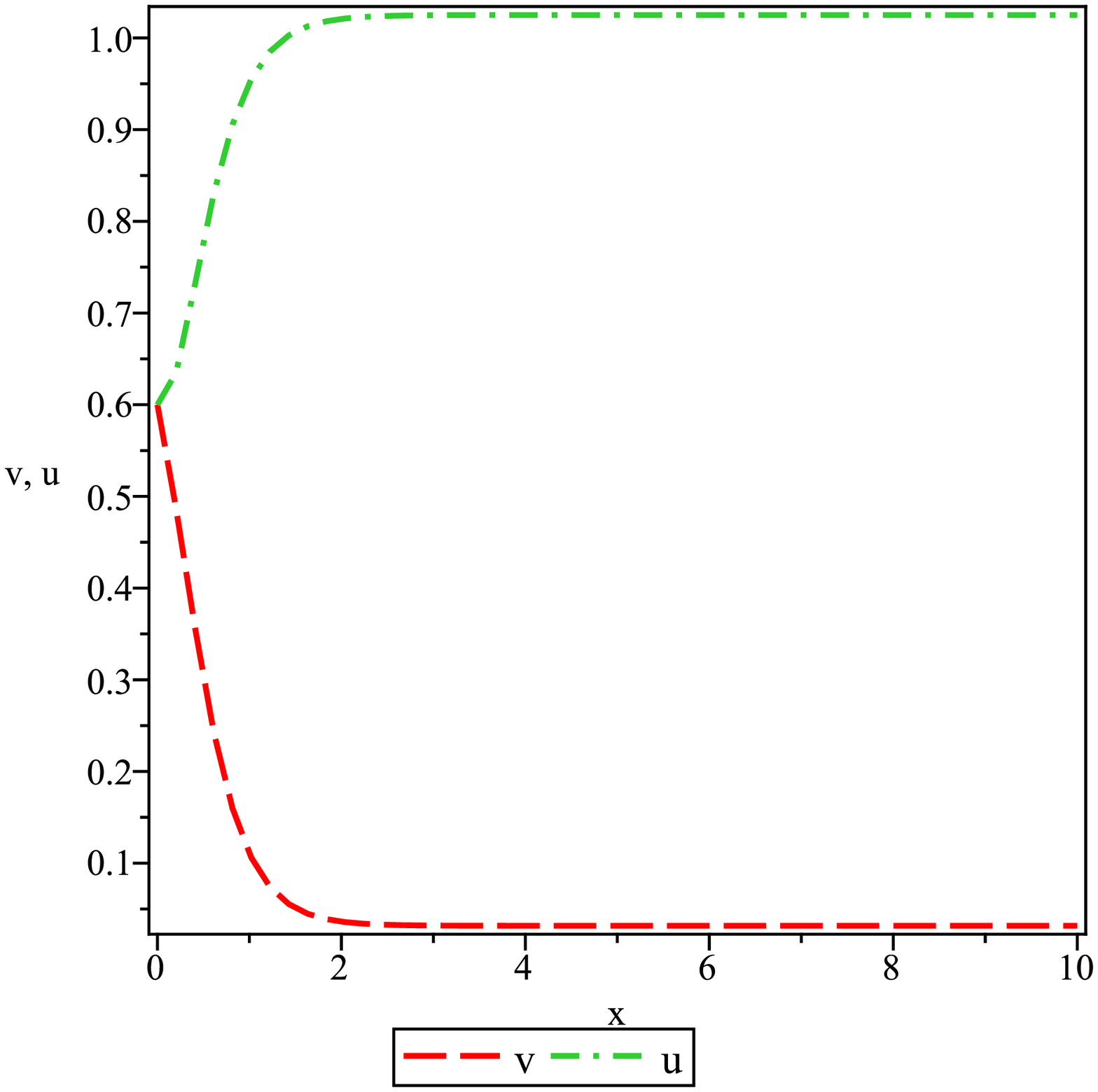}\\
\caption{The dimensionless density parameters are plotted against
e-folding time. The initial condition is $v(0)=0.6$, $u(0)=0.6$.
Other parameters are fixed at $b=0.03$, $A=0.3$, $B=0.5$ and
$\alpha=0.5$.}
\end{figure}
\begin{figure}
\includegraphics[scale=.4]{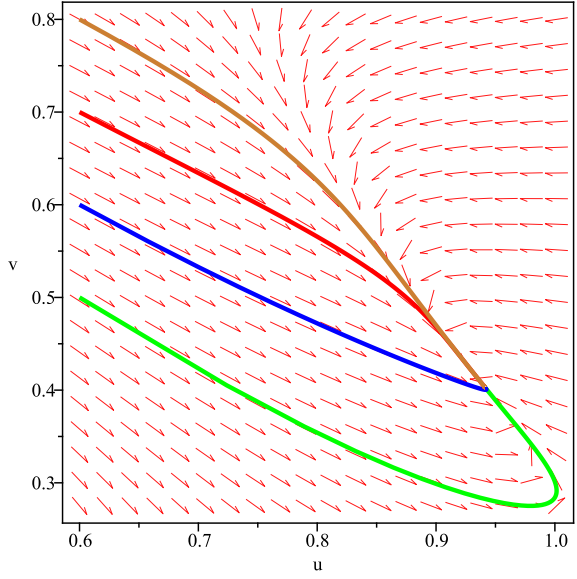}\\
\caption{The phase space diagram of parameters depicting an
attractor solution. The initial conditions chosen are $v(0)=0.5$,
$u(0)=0.6$ (green); $v(0)=0.6$, $u(0)=0.6$ (blue); $v(0)=0.7$,
$u(0)=0.6$ (red); $v(0)=0.8$, $u(0)=0.6$ (brown). Other parameters
are fixed at $b=0.3$, $A=0.3$, $B=0.5$ and $\alpha=1$.}
\end{figure}
\begin{figure}
\includegraphics[scale=.4]{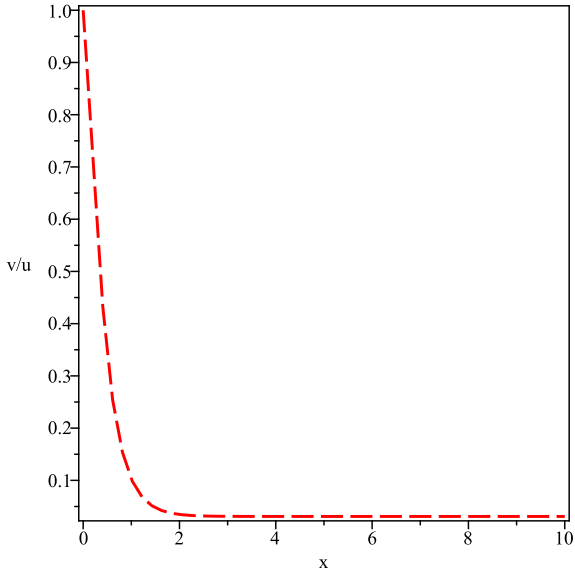}\\
\caption{The ratio of density parameters is shown against e-folding
time.. The initial condition chosen is $v(0)=0.6$, $u(0)=0.6$. Other
parameters are fixed at $b=0.03$, $A=0.3$, $B=0.5$ and
$\alpha=0.5$.}
\end{figure}
\begin{figure}
\includegraphics[scale=.4]{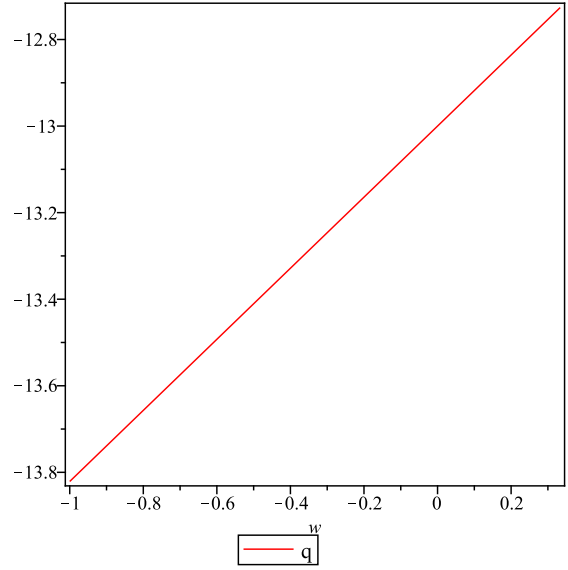}\\
\caption{The deceleration parameter is plotted against the state
parameter. Other parameters are fixed at $b=0.3$, $\epsilon=0.4$,
$A=0.4$ and $B=0.5$.}
\end{figure}
\begin{figure}
\includegraphics[scale=.4]{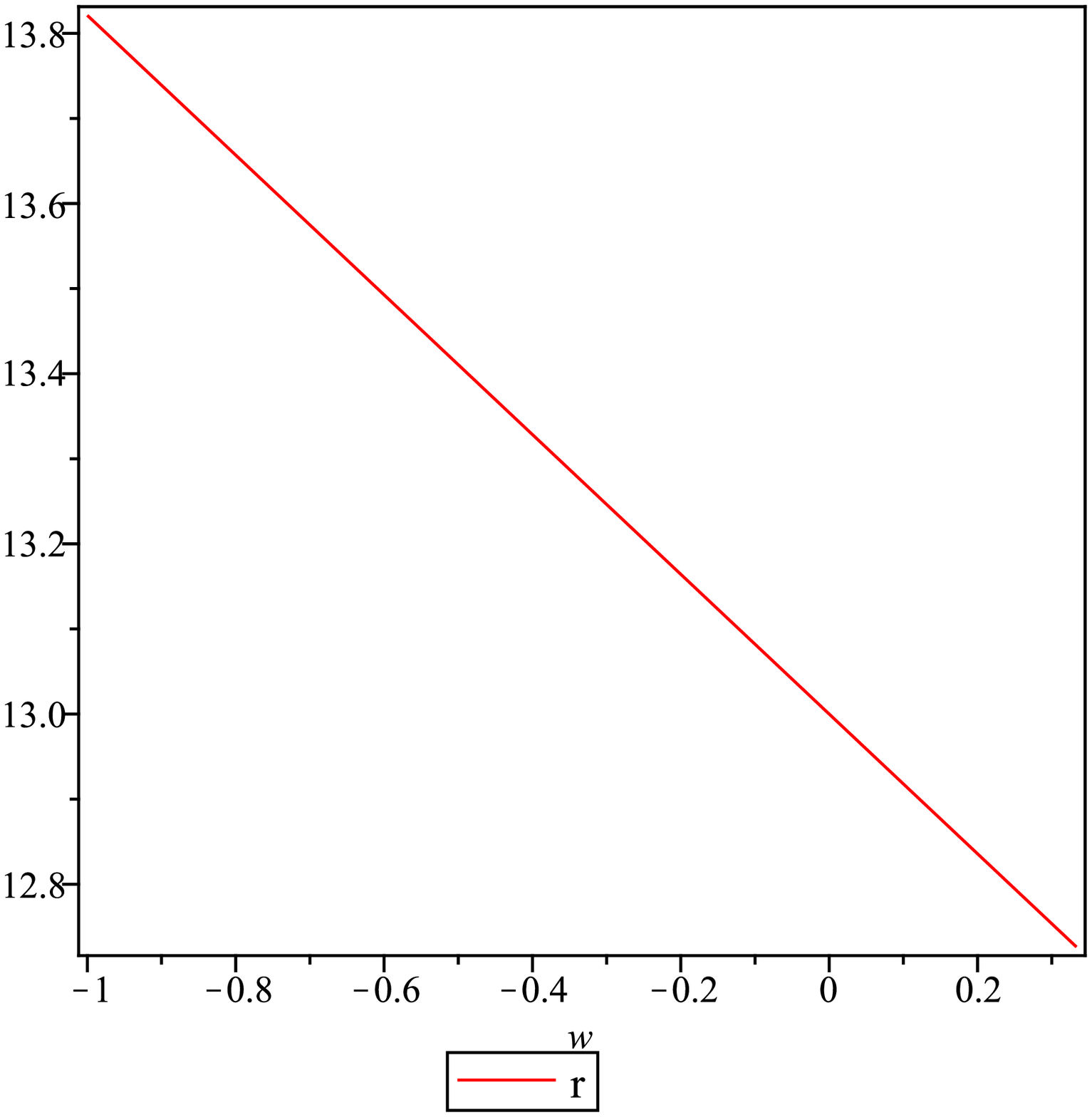}\\
\caption{The statefinder parameter $r$ is plotted against the state
parameter. Other parameters are fixed at $b=0.3$, $\epsilon=0.4$,
$A=0.4$ and $B=0.5$.}
\end{figure}
\begin{figure}
\includegraphics[scale=.4]{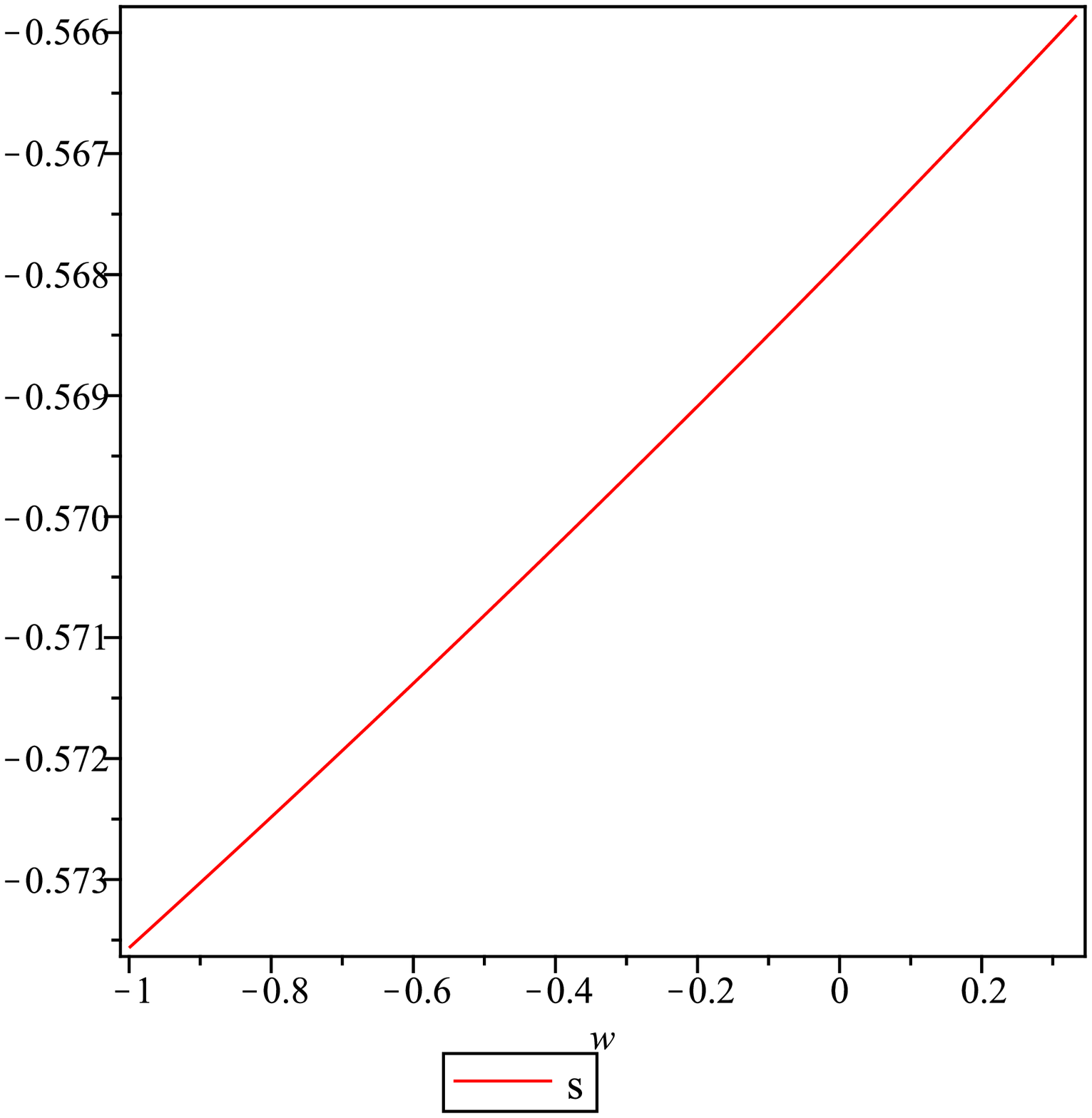}\\
\caption{The statefinder parameter $s$ is plotted against the state
parameter. Other parameters are fixed at $b=0.3$, $\epsilon=0.4$,
$A=0.4$ and $B=0.5$.}
\end{figure}


\begin{thebibliography}{}
\bibitem[\protect\citeauthoryear{Perlmutter et al}{1999}]{perl} Perlmutter S. et al., 1999, Astrophys. J., 517, 565
\bibitem[\protect\citeauthoryear{Riess et al}{1998}]{ries} Riess A. et al., 1998 Astron. J., 116, 1009
\bibitem[\protect\citeauthoryear{Sahni \& Starobinsky}{2000}]{sahni} Sahni V. \& Starobinsky A.A., 2000 Int. J. Mod. Phys. D 9, 373
\bibitem[\protect\citeauthoryear{Padmanabhan}{2003}]{paddy} Padmanabhan T., 2003 Phys. Rept. 380, 235
\bibitem[\protect\citeauthoryear{Copeland et al}{2006}]{cope} Copeland E.J., Sami M., \& Tsujikawa T., 2006 Int. J. Mod. Phys D., 15, 1753
\bibitem[\protect\citeauthoryear{Peebles \& Ratra}{2006}]{pee} Peebles P.J.E., \& Ratra B., 1988 Astrophys. J. Lett., 325, L17
\bibitem[\protect\citeauthoryear{Jamil et al}{2010}]{jamil1} Jamil M., Saridakis E.N., \& Setare M.R., 2010 Phys. Rev. D 81, 023007
\bibitem[\protect\citeauthoryear{Jamil \& Saridakis}{2010}]{jamil2} Jamil M. \& Saridakis E.N., 2010 JCAP 07, 028
\bibitem[\protect\citeauthoryear{Jamil \& Farooq}{2010}]{jamil3} Jamil M. \& Farooq M.U., 2010 JCAP 03, 001
\bibitem[\protect\citeauthoryear{Jamil et al}{2010}]{jamil4} Jamil M., Sheykhi A., \& Farooq M.U., 2010 Int. J. Mod. Phys. D 19, 1831
\bibitem[\protect\citeauthoryear{Jamil \& Rahman}{2009}]{jamil5} Jamil M., \& Rahman F., 2009 Eur. Phys. J. C 64, 97
\bibitem[\protect\citeauthoryear{Setare}{2007}]{setare1} Setare M.R., 2007 Eur. Phys. J. C 50, 991
\bibitem[\protect\citeauthoryear{Setare}{2006}]{setare2} Setare M.R., 2006 Phys. Lett. B 642, 1
\bibitem[\protect\citeauthoryear{Hu \& Ling}{2006}]{hu} Hu B., \& Ling Y., 2006 Phys. Rev. D 73, 123510
\bibitem[\protect\citeauthoryear{Wu \& Yu}{2007}]{wu} Wu P., \& Yu H., 2007 Class. Quantum Grav. 24, 4661
\bibitem[\protect\citeauthoryear{Jamil}{2010}]{jamil6} Jamil M., 2010 Int. J. Theor. Phys. 49, 62
\bibitem[\protect\citeauthoryear{Setare}{2006}]{setare67} Setare M.R., 2006, Phys. Lett. B 642, 1
\bibitem[\protect\citeauthoryear{Wu \& Yu}{2008}]{wu1} Wu P., \& Zhang S.N., 2008 JCAP 06, 007
\bibitem[\protect\citeauthoryear{Chen et al}{2008}]{chen} Chen S., Wang B., \& Jing J., 2008 Phys. Rev. D 78, 123503
\bibitem[\protect\citeauthoryear{Fu et al}{2008}]{fu} Fu X., Yu H., \& Wu P., 2008 Phys. Rev. D 78, 063001
\bibitem[\protect\citeauthoryear{Guo \& Zhang}{2005}]{guo} Guo Z-K., \& Zhang Y-Z., 2005 Phys. Rev. D 71, 023501
\bibitem[\protect\citeauthoryear{Campo et al}{2008}]{campo22} del Campo S. et al., arXiv:0812.2210v1 [gr-qc]
\bibitem[\protect\citeauthoryear{Alcaniz \& Lima}{2005}]{lima} Alcaniz J.S. \& Lima J.A.S., 2005 Phys. Rev. D 72, 063516
\bibitem[\protect\citeauthoryear{Feng et al}{2008}]{feng} Feng C. et al., 2008 Phys. Lett. B, 665, 111
\bibitem[\protect\citeauthoryear{Lu et al}{2008}]{lu} Lu J. et al., 2008 Phys. Lett. B 662, 87
\bibitem[\protect\citeauthoryear{Jun \& Zhou}{2005}]{jun} Dao-Jun L., \& Xin-Zhou L., 2005 Chin. Phys. Lett., 22, 1600
\bibitem[\protect\citeauthoryear{Jing et al}{2008}]{jing} Jing H. et al., 2008 Chin. Phys. Lett., 25, 347
\bibitem[\protect\citeauthoryear{Debnath}{2007}]{debnath} Debnath U., arXiv:0710.1708 [gr-qc]
\bibitem[\protect\citeauthoryear{Barreiro \& Sen}{2004}]{sen} Barreiro T., \& Sen A.A., 2004 Phys. Rev. D 70, 124013
\bibitem[\protect\citeauthoryear{Carturan \& Finelli}{2003}]{carturan} Carturan D. \& Finelli F., 2003 Phys. Rev. D 68, 103501
\bibitem[\protect\citeauthoryear{Barriero et al}{2008}]{barriero} Barriero T. et al., 2008 Phys. Rev. D 78, 043530
\bibitem[\protect\citeauthoryear{Makler et al}{2003}]{makler} Makler M. et al., 2003 Phys. Lett. B 555, 1
\bibitem[\protect\citeauthoryear{Setare}{2009}]{setarea} Setare M.R., 2009 Int. J. Mod. Phys. D 18, 419
\bibitem[\protect\citeauthoryear{Setare}{2007}]{setareb} Setare M.R., 2007 Phys. Lett. B 648, 329
\bibitem[\protect\citeauthoryear{Setare}{a2007}]{setarec} Setare M.R., 2007 Eur. Phys. J. C 52, 689
\bibitem[\protect\citeauthoryear{Kamenshchik}{2001}]{kamenshchik} Kamenshchik A. et al., 2001 Phys. Lett. B 511, 265
\bibitem[\protect\citeauthoryear{Panotopoulos}{2008}]{grigoris} Panotopoulos G., 2008 Phys. Rev. D 77, 107303
\bibitem[\protect\citeauthoryear{Sahni et al}{2003}]{sahni1} Sahni V. et al., 2003 JETP 77, 201
\end{thebibliography}
\end{document}